\documentclass[a4paper]{jpconf-kb}
\usepackage{amsmath,amsthm}
\usepackage{iopams}
\usepackage{natbib}
\usepackage{graphicx}

\usepackage{multind}
\makeindex{subject}
\makeindex{authors}

\newtheorem{thm}{Theorem}

\newtheorem{lem}[thm]{Lemma}

\theoremstyle{definition}

\begin{document}
\graphicspath{{FIGURES/}}

\title{A complete topological invariant for braided magnetic fields}

\author{A R Yeates$^1$ and G Hornig$^2$}
\address{$^1$ Department of Mathematical Sciences, Durham University, Durham, DH1 3LE, UK}
\address{$^2$ Division of Mathematics, University of Dundee, Dundee, DD1 4HN, UK}

\ead{\mailto{anthony.yeates@durham.ac.uk}, \mailto{gunnar@maths.dundee.ac.uk}}

\index{authors}{Yeates, A.R.} \index{authors}{Hornig, G.}

\begin{abstract}
A topological flux function is introduced to quantify the topology of magnetic braids: non-zero line-tied magnetic fields whose field lines all connect between two boundaries. This scalar function is an ideal invariant defined on a cross-section of the magnetic field, whose integral over the cross-section yields the relative magnetic helicity. Recognising that the topological flux function is an action in the Hamiltonian formulation of the field line equations, a simple formula for its differential is obtained. We use this to prove that the topological flux function uniquely characterises the field line mapping and hence the magnetic topology. A simple example is presented.
\end{abstract}

\section{Introduction}

In this paper we present a complete topological invariant\index{subject}{invariant} for so-called \emph{magnetic braids}\index{subject}{magnetic braid}: magnetic fields in a flux tube with monotone guide field. Such magnetic fields arise, for example in coronal loops in the Sun's atmosphere \citep{reale2010},\index{authors}{Reale, F.} or in toroidal fusion devices \citep{morrison2000}\index{authors}{Morrison, P.J.}. In both cases, the magnetic field is embedded in a highly-conducting plasma so that the magnetic topology---or linking and connectivity of magnetic field lines\index{subject}{magnetic field lines}---is approximately preserved. As such, topological invariants typically play a significant role in the dynamics of the plasma and its magnetic field \citep{woltjer1958,taylor1986,brown1999,yeates2010,candelaresi2011}.
\index{authors}{Woltjer, L.}
\index{authors}{Taylor, J.B.}
\index{authors}{Brown, M.R.}
\index{authors}{Canfield, R.C.}
\index{authors}{Pevtsov, A.A.}
\index{authors}{Yeates, A.R.}
\index{authors}{Hornig, G.}
\index{authors}{Wilmot-Smith, A.L.}
\index{authors}{Candelaresi, S.}
\index{authors}{Brandenburg, A.}

\begin{figure}[!ht]
\begin{minipage}{0.45\textwidth}
\includegraphics[width=0.7\textwidth]{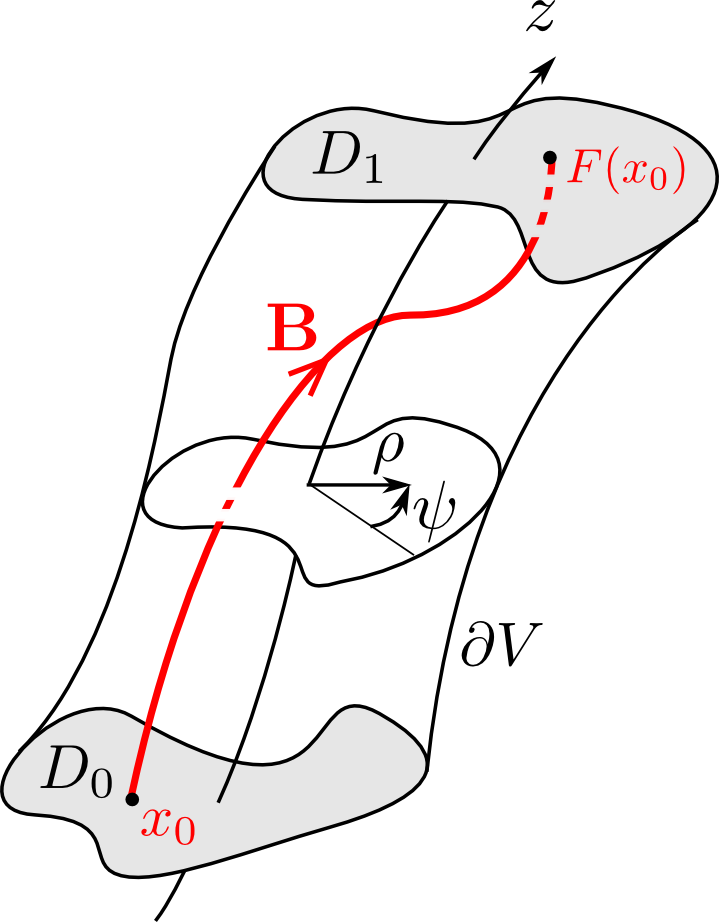}
\caption{\label{fig:domain}Coordinates and notation for a magnetic braid ${\bf B}$ in a flux tube $V$, with field line mapping $F:D_0\rightarrow D_1$.}
\end{minipage}\hspace{0.1\textwidth}%
\begin{minipage}{0.45\textwidth}
\includegraphics[width=0.7\textwidth]{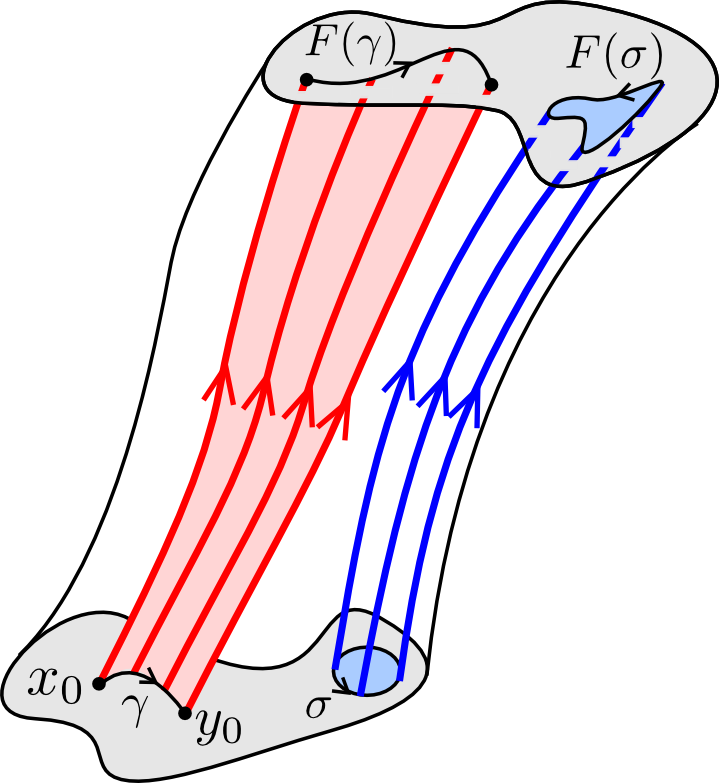}
\caption{\label{fig:curves}The physical meaning of Lemma \ref{lem:da} may be understood by integrating on the lower boundary $D_0$ along an open curve $\gamma$ or a closed curve $\sigma$.}
\end{minipage} 
\end{figure}

Figure \ref{fig:domain} shows our simply connected ``flux tube'' $V$, with a set $(\rho,\psi,z)$ of orthogonal curvilinear coordinates satisfying $\nabla z\cdot(\nabla\rho\times\nabla\psi)\neq 0$. The magnetic field ${\bf B}$ is assumed to satisfy $B_z>0$ everywhere in $V$, so that $\nabla z$ aligns with the ``axis'' of the flux tube. We take $\psi$ to be an angular coordinate and $\rho$ to be a radial coordinate in each plane of constant $z$. On the side boundaries of the tube, we require ${\bf B}\cdot{\bf n}=0$, so that all magnetic field lines stretch from one end $D_0$ of the flux tube to the other end $D_1$. We can then define the \emph{field line mapping}\index{subject}{field line mapping} $F:D_0\rightarrow D_1$ by $F(x_0)=f(x_0;1)$, where $f(x_0;z)$ denotes the magnetic field line rooted at $x_0\in D_0$.

Two magnetic braids ${\bf B}$, $\widetilde{\bf B}$ are \emph{topologically equivalent} if and only if one can be reached from the other by an ideal evolution
\begin{equation}
\frac{\partial{\bf B}}{\partial t} = \nabla\times({\bf v}\times{\bf B})
\end{equation}
 with ${\bf v}|_{\partial V}=0$ throughout. We shall assume that their respective field line mappings match on the side boundary, $F|_{\partial D_0}=\widetilde{F}|_{\partial D_0}$, with the same winding number. In that case, ${\bf B}$ and $\widetilde{\bf B}$ are topologically equivalent if and only if $F=\widetilde{F}$. Notice that $F$ has two components. Our main result is to prove that a single scalar function, which we call a ``topological flux function''\index{subject}{topological flux function} is both necessary and sufficient to determine the topology. This we state in the following theorem.
\begin{thm}
Let ${\bf B}$, $\widetilde{\bf B}$ be two magnetic braids on $V$, with respective field line mappings $F$, $\widetilde{F}$ that agree on $\partial D_0$, with the same winding number. Let ${\cal A}$, $\widetilde{\cal A}$ be their respective topological flux functions with the same reference field ${\bf A}^{\rm ref}$ such that $A_\rho^{\rm ref}=0$. Then ${\cal A}=\widetilde{\cal A}$ if and only if $F=\widetilde{F}$.
\label{thm:main}
\end{thm}
It is well-known \citep{berger1984,brown1999}
\index{authors}{Berger, M.A.}%
\index{authors}{Field, G.B.}%
\index{authors}{Brown, M.R.}%
\index{authors}{Canfield, R.C.}%
\index{authors}{Pevtsov, A.A.}%
that having the same total magnetic helicity \index{subject}{helicity}$H_r$ (which will be defined below) is a necessary condition for topological equivalence but not a sufficient one (two magnetic braids can have the same $H_r$ but different $F$). Indeed there are an infinity of other ideal invariants.\index{subject}{invariant} For example, if $\theta_{x_0,y_0}(z)$ denotes the orientation of the line at height $z$ between a pair of field lines $f(x_0;z)$ and $f(y_0;z)$, then their pairwise linking number \index{subject}{linking number}
\begin{equation}
c_{x_0,x_1} = \frac{1}{2\pi}\int_0^1\frac{\mathrm{d}\theta_{x_0,y_0}(z)}{\mathrm{d}z}\,\mathrm{d}z
\end{equation}
is an invariant \citep{berger1986,berger2006}.
\index{authors}{Berger, M.A.}%
\index{authors}{Prior, C.}%
 So are analogous higher-order measures of the linking between triplets, 4-tuplets, etc., of field lines. Theorem \ref{thm:main} asserts that all of these ideal invariants are contained in ${\cal A}$ because, mathematically speaking, it is a \emph{complete} invariant. It is the most economical description of field topology equivalence classes that is sought by \citet{berger1986}.
\index{authors}{Berger, M.A.} %
  In future, the topological flux function should provide a useful tool not just to recognise but also to \emph{quantify} changes in magnetic topology during non-ideal evolutions. For an initial application to measuring magnetic reconnection, see \citet{yeates2011}.
\index{authors}{Yeates, A.R.}%
\index{authors}{Hornig, G.}%

For the alternative geometry of a half-space, \citet{berger1988}
\index{authors}{Berger, M.A.}%
 introduces ${\cal A}$ as the helicity of an infinitesimal flux tube around a single field line. Indeed, \citet{taylor1986}
 \index{authors}{Taylor, J.B.}%
  points out that each closed field line yields such an invariant in a perfectly-conducting plasma. This paper aims to interpret these constraints in a different light. We show that Theorem \ref{thm:main} is a consequence of ${\cal A}$ being the {\it action} the Hamiltonian system\index{subject}{Hamiltonian system} of the magnetic field lines \citep{cary1983}.
  \index{authors}{Cary, J.R.}%
  \index{authors}{Littlejohn, R.G.}%
   The Hamiltonian theory suggests an elegant expression for ${\cal A}$ in terms of differential forms that we exploit.

A less general form of Theorem \ref{thm:main} has recently appeared in \citet{yeates2012}.
\index{authors}{Yeates, A.R.}%
\index{authors}{Hornig, G.}%
 Here we present the results for a more general shape of flux tube with more general boundary conditions, and elaborate further on the Hamiltonian theory. We begin in Section \ref{sec:def} with the basic definition of ${\cal A}$, before discussing the Hamiltonian interpretation in Section \ref{sec:ham}. The proof of Theorem \ref{thm:main} is given in Section \ref{sec:proof} and two explicit examples in Section \ref{sec:ex}.

\section{Topological Flux Function} \label{sec:def}

We define the \emph{topological flux function} \index{subject}{topological flux function}${\cal A}:D_0\rightarrow\mathbb{R}$ as the line integral of ${\bf A}$ along the magnetic field line rooted at $x_0\in D_0$:
\begin{equation}
{\cal A}(x_0) = \int_{x_0}^{F(x_0)}{\bf A}\big(f(x_0;z)\big)\cdot\,\mathrm{d}{\bf l}.
\end{equation}
As it stands, this definition is gauge dependent. Under a gauge transformation \index{subject}{gauge transformation}${\bf A}\rightarrow{\bf A}+\nabla\chi$, it is easy to see that ${\cal A} \rightarrow {\cal A} + F^*\chi - \chi$, where the pull-back notation means $\Big(F^*\chi\Big)(x_0)=\chi\big(F(x_0)\big)$. We impose the gauge condition
\begin{equation}
{\bf n}\times{\bf A}|_{\partial V} = {\bf n}\times{\bf A}^{\rm ref}|_{\partial V},
\label{eqn:gauge}
\end{equation}
where ${\bf A}^{\rm ref}$ is the vector potential of a reference field ${\bf B}^{\rm ref}$ that matches ${\bf B}\cdot{\bf n}$ on $\partial V$ but is otherwise arbitrary. This particular choice of gauge condition is motivated by the relative magnetic helicity (Section \ref{sec:hr}).\index{subject}{helicity} Although this still leaves some freedom in $\chi$, and consequently in ${\cal A}$, it is enough to make ${\cal A}$ an ideal invariant.\index{subject}{invariant} To see this, assume an ideal evolution
\begin{equation}
\frac{\partial{\bf A}}{\partial t} = {\bf v}\times{\bf B} + \nabla\Phi.
\end{equation}
The boundary conditions of ${\bf v}|_{\partial V}=0$ and fixed ${\bf n}\times{\bf A}|_{\partial V}$ mean that ${\bf n}\times\nabla\Phi|_{\partial V}=0$. Then, using the formula for the rate of change of a line integral over a moving domain \citep[e.g.,][]{frankel1997},
\index{authors}{Frankel, T.}%
 we find
\begin{align}
\frac{\mathrm{d}{\cal A}}{\mathrm{d}t}&=\frac{\mathrm{d}}{\mathrm{d}t}\int_{x_0}^{F(x_0)}{\bf A}\cdot\,\mathrm{d}{\bf l}\\
&=\int_{x_0}^{F(x_0)}\left(\frac{\partial{\bf A}}{\partial t} - {\bf v}\times\nabla\times{\bf A} + \nabla({\bf v}\cdot{\bf A}) \right)\cdot\,\mathrm{d}{\bf l}\\
&=\int_{x_0}^{F(x_0)}\nabla(\Phi + {\bf v}\cdot{\bf A})\cdot\,\mathrm{d}{\bf l} = 0.
\end{align}

\subsection{Relation to Magnetic Helicity} \label{sec:hr}

\citet{berger1988} 
\index{authors}{Berger, M.A.}%
defines ${\cal A}$ as a ``field line helicity'',\index{subject}{helicity!field line} or limiting magnetic helicity\index{subject}{helicity} in an infinitesimal flux tube around a single magnetic field line. Since our domain is magnetically open, we define a gauge invariant relative helicity\index{subject}{helicity!relative} \citep{berger1984,finn1985}
\index{authors}{Berger, M.A.}%
\index{authors}{Field, G.B.}%
\index{authors}{Finn, J.H.}%
\index{authors}{Antonsen, T.M.}%
 relative to the reference field ${\bf B}^{\rm ref}=\nabla\times{\bf A}^{\rm ref}$. Assuming (\ref{eqn:gauge}), the relative helicity may be written
\begin{equation}
H_r = \int_V({\bf A} + {\bf A}^{\rm ref})\cdot({\bf B} - {\bf B}^{\rm ref})\,\mathrm{d}^3x = \int_V{\bf A}\cdot{\bf B}\,d^3x - \int_V{\bf A}^{\rm ref}\cdot{\bf B}^{\rm ref}\,\mathrm{d}^3x.
\end{equation}
The last term is a constant (call it $H^{\rm ref}$) which depends only on the choice of reference field. For the first term, let $x_0$ be the footpoint of the field line through $x=(\rho,\psi,z)$, so that $x=f(x_0;z)$. Changing coordinates gives
\begin{align}
H_r - H^{\rm ref} &= \int_V{\bf A}(x)\cdot{\bf B}(x)\,\mathrm{d}^3x,\\
&= \int_V{\bf A}\big(f(x_0;z)\big)\cdot{\bf B}\big(f(x_0;z)\big)\frac{B_z(x_0)}{B_z\big(f(x_0;z)\big)}\mathrm{d}^2x_0\mathrm{d}z,\\
&= \int_{D_0}{\cal A}(x_0)B_z(x_0)\,\mathrm{d}^2x_0.
\end{align}
Thus we see that ${\cal A}$ represents the helicity per field line.

\section{Hamiltonian Interpretation} \label{sec:ham}

We will show in this section that the differential of ${\cal A}$ has the following succinct formula in terms of differential forms \citep{frankel1997},
\index{authors}{Frankel, T.}%
 where $h_\rho\equiv||\partial({\bf r})/\partial\rho||$, $h_\psi\equiv||\partial({\bf r})/\partial\psi||$ are the coordinate scale factors.
\begin{lem}
Let ${\bf B}$ be a magnetic braid with field line mapping $F$ and topological flux function ${\cal A}$. Then
\[
\mathrm{d}{\cal A} = F^*\alpha - \alpha,
\]
where $\alpha=A_\rho^{\rm ref}h_\rho\,\mathrm{d}\rho + A_\psi^{\rm ref}h_\psi\,\mathrm{d}\psi$ is the 1-form associated to ${\bf A}^{\rm ref}$ on $D_0$ and (perhaps differently) on $D_1$.
\label{lem:da}
\end{lem}
This formula will be used to prove Theorem \ref{thm:main} in Section \ref{sec:proof}. It may be understood physically by integrating along a curve $\gamma\in D_0$, so that
\begin{equation}
\int_\gamma \mathrm{d}{\cal A} = \int_{\gamma}F^*\alpha - \int_\gamma\alpha.
\label{eqn:intdA}
\end{equation}
In vector notation, equation \eqref{eqn:intdA} reads
\begin{equation}
{\cal A}(y_0) - {\cal A}(x_0) = \int_{F(\gamma)}{\bf A}\cdot\,\mathrm{d}{\bf l} - \int_\gamma{\bf A}\cdot\,\mathrm{d}{\bf l},
\label{eqn:gamma}
\end{equation}
where $x_0, y_0\in D_0$ are the start and end points of $\gamma$ (see Figure \ref{fig:curves}). If $y_0\neq x_0$ then $\gamma$ is an open curve, and (\ref{eqn:gamma}) expresses the fact that the field lines rooted in $\gamma$ form a flux surface with $\oint{\bf A}\cdot\,\mathrm{d}{\bf l}=0$. This is the vertical shaded surface in Figure \ref{fig:curves}.  On the other hand, if $y_0=x_0$ then the curve is closed (the curve $\sigma$ in Figure \ref{fig:curves}), in which case (\ref{eqn:gamma}) expresses conservation of (vertical) magnetic flux in the corresponding flux tube. 

\subsection{$\cal A$ as an Action}

It is insightful to derive Lemma \ref{lem:da} by interpreting ${\cal A}$  as an action in a Hamiltonian system.\index{subject}{Hamiltonian system} Suppose we parametrise a magnetic field line in terms of its length $l$ as ${\bf x}(l)$, and think of the topological flux function as a functional,
\begin{equation}
{\cal A}(x_0) = \int_{x_0}^{F(x_0)}{\bf A}({\bf x})\cdot\frac{\mathrm{d}{\bf x}}{\mathrm{d}l}\,\mathrm{d}l.
\end{equation}
Then \citet{cary1983}
\index{authors}{Cary, J.R.}%
\index{authors}{Littlejohn, R.G.}%
 point out that extremising ${\cal A}$ for given ${\bf A}$ and fixed end-points gives the path of the field line ${\bf x}(l)$ through the domain. To see this, note that the Euler-Lagrange equations are
\begin{equation}
\frac{\partial L}{\partial x_j} - \frac{\mathrm{d}}{\mathrm{d}l}\left(\frac{\partial L}{\partial(\partial x_j/\partial l)}\right)=0.
\end{equation}
In our case, the Lagrangian is $L={\bf A}\cdot({\mathrm{d}{\bf x}}/{\mathrm{d}l})$, so
\begin{equation}
\frac{\partial A_i}{\partial x_j}\frac{\partial x_i}{\partial l} - \frac{\partial x_i}{\partial l}\frac{\partial A_j}{\partial x_i} = 0,
\end{equation}
which is equivalent to
\begin{equation}
(\nabla\times{\bf A})\times\frac{\mathrm{d}{\bf x}}{\mathrm{d}l} = 0.
\end{equation}
Hence $\mathrm{d}{\bf x}/\mathrm{d}l$ is everywhere parallel to ${\bf B}$, and ${\bf x}(l)$ is a magnetic field line.

In fact, it is well known that the equations of the magnetic field lines are a Hamiltonian system. To show this, we follow \citet{cary1983}
 \index{authors}{Cary, J.R.}
\index{authors}{Littlejohn, R.G.}%
and change the gauge of ${\bf A}$ so as to write the action explicitly in the canonical form of a Hamiltonian system. A gauge transformation ${\bf A}\rightarrow{\bf A}+\nabla\chi$ changes the integrand (the Lagrangian) to $L + \mathrm{d}\chi/\mathrm{d}l$ and also changes the integral (the action ${\cal A}$), but leaves the Euler-Lagrange equations unchanged. If we set $A_\rho=0$ everywhere in space, then ${\bf A}\cdot\,\mathrm{d}{\bf x}=A_\psi h_\psi\,\mathrm{d}\psi + A_zh_z\,\mathrm{d}z$. Making the identifications
$p\leftrightarrow A_\psi(\rho,\psi,z)h_\psi,  q\leftrightarrow \psi,  t\leftrightarrow z,  H \leftrightarrow -A_z(\rho,\psi,z)h_z$,
our action becomes
\begin{equation}
{\cal A} = \int_{x_0}^{F(x_0)}\left(p\frac{\mathrm{d}q}{\mathrm{d}t} - H(p,q,t)\right)\,\mathrm{d}t,
\label{eqn:ham}
\end{equation}
which is a 1 degree-of-freedom Hamiltonian system in canonical form. The generalised coordinate is $\psi$, the generalised momentum is $A_\psi h_\psi$, and the Hamiltonian is $-A_z h_z$. Time is the $z$-direction, so our Hamiltonian is, in general, time dependent.

The gauge where $A_\rho=0$ everywhere may be called a \emph{canonical gauge} since it renders the Hamiltonian system in canonical form. For Theorem \ref{thm:main} it is sufficient that ${\bf n}\times{\bf A}|_{\partial V}$ be in canonical gauge, so in practice we restrict the gauge of ${\bf A}^{\rm ref}$ to be canonical, i.e. $A_\rho^{\rm ref}=0$. The gauge $A_\rho=0$ is equivalent to writing ${\bf B}$ in the form ${\bf B}=\nabla(A_\psi h_\psi)\times\nabla\psi + \nabla(A_z h_z)\times\nabla z$ \citep{boozer1983,yoshida1994}.
\index{authors}{Boozer, A.H.}%
\index{authors}{Yoshida, Z.}%

\subsection{Proof of Lemma \ref{lem:da}}

Given that ${\cal A}$ is the action in a Hamiltonian system, we can use known properties of Hamiltonian systems. Let $f_z$ be the ``height-$z$'' mapping of our field line system, defined by $f_z(x_0)=f(x_0;z)$. This mapping must preserve magnetic flux. In terms of differential forms,
$f_z^*\omega - \omega = 0$,
where $\omega = B_zh_\rho h_\psi\,\mathrm{d}\rho\wedge \mathrm{d}\psi$. Every Hamiltonian system preserves such a symplectic form $\omega$. But notice that
\begin{equation}
\omega = \left(\frac{\partial}{\partial \rho}\big(A_\psi h_\psi\big) - \frac{\partial}{\partial \psi}\big(A_\rho h_\rho \big) \right)\,\mathrm{d}\rho\wedge \mathrm{d}\psi = \mathrm{d}\alpha
\end{equation}
where $\alpha=A_\psi h_\psi\,\mathrm{d}\psi + A_\rho h_\rho\,\mathrm{d}\rho$. The primitive $\alpha$ is called the canonical/Liouville 1-form.

Assume now that ${\bf A}$ is in canonical gauge as above, so that $\alpha=A_\psi h_\psi\,\mathrm{d}\psi$. Then Lemma \ref{lem:da} follows from the following argument \citep{haro1998}.
\index{authors}{Haro, A.}%
 In canonical coordinates, Hamilton's equations have the form
\begin{equation}
\dot{p}=-\frac{\partial H}{\partial q},\quad \dot{q}=\frac{\partial H}{\partial p}.
\end{equation}
If we define the Hamiltonian vector field ${\bf x}(t)\equiv(x^q,x^p)= (\dot{q},\dot{p})$, then Hamilton's equations may be succinctly written \citep{marsden1994}
\index{authors}{Marsden, J.E.}%
\index{authors}{Ratiu, T.S.}%
in terms of $\alpha$ as
\begin{equation}
i_{\bf x}\,\mathrm{d}\alpha = -\mathrm{d}H.
\end{equation}
By the definition of the Lie derivative \citep[Theorem 4.3.1 of ][]{marsden1994}, using Cartan's magic formula, and the fact that differentials and pull-backs commute, we have
\begin{equation}
\frac{d}{ds}(f_s^*\alpha) = f_s^*{\cal L}_{{\bf x}(s)}\alpha
=f_s^*\big(i_{{\bf x}(s)}\,\mathrm{d}\alpha +\mathrm{d}i_{{\bf x}(s)}\alpha\big)
=\mathrm{d}f_s^*\big(i_{{\bf x}(s)}\alpha - H \big).
\end{equation}
Integrating both sides from $s=0$ to $s=z$ yields the desired formula
\begin{equation}
f_z^*\alpha-\alpha = \mathrm{d}\Big(\int_0^zf_s^*\big(i_{{\bf x}(s)}\alpha - H \big)\,\mathrm{d}s \Big).
\end{equation}
Taking $z=1$ in this expression and recognising (\ref{eqn:ham}) gives
\begin{equation}
F^*\alpha-\alpha = \mathrm{d}\Big(\int_0^1f_z^*\big(p\dot{q} - H \big)\,\mathrm{d}z \Big)
= \mathrm{d}\Big(\int_0^1\big(p\dot{q} - H \big)\circ f_z\,\mathrm{d}z \Big)=\mathrm{d}{\cal A},
\end{equation}
which is Lemma \ref{lem:da}.

To show that Lemma \ref{lem:da} holds even in non-canonical gauge, suppose that $F^*\alpha-\alpha=\mathrm{d}{\cal A}$ for some ${\bf A}$. Under the change of gauge ${\bf A}\rightarrow {\bf A}+\nabla\chi$, we find $\alpha\rightarrow\alpha + \mathrm{d}\chi$, so
\begin{equation}
F^*\alpha-\alpha \quad \rightarrow\quad F^*\alpha - \alpha + \mathrm{d}(F^*\chi-\chi)=\mathrm{d}(\cal A + F^*\chi-\chi).
\end{equation}
But this is precisely $\mathrm{d}{\cal A}$ in the new gauge (Section \ref{sec:def}), so if the Lemma holds in one gauge it holds in any gauge.\index{subject}{gauge transformation}

\section{Proof of Theorem \ref{thm:main}} \label{sec:proof}

Lemma \ref{lem:da} allows for an elegant proof of Theorem \ref{thm:main}.

\subsection*{Necessity}
\begin{figure}[h]
\includegraphics[width=0.3\textwidth]{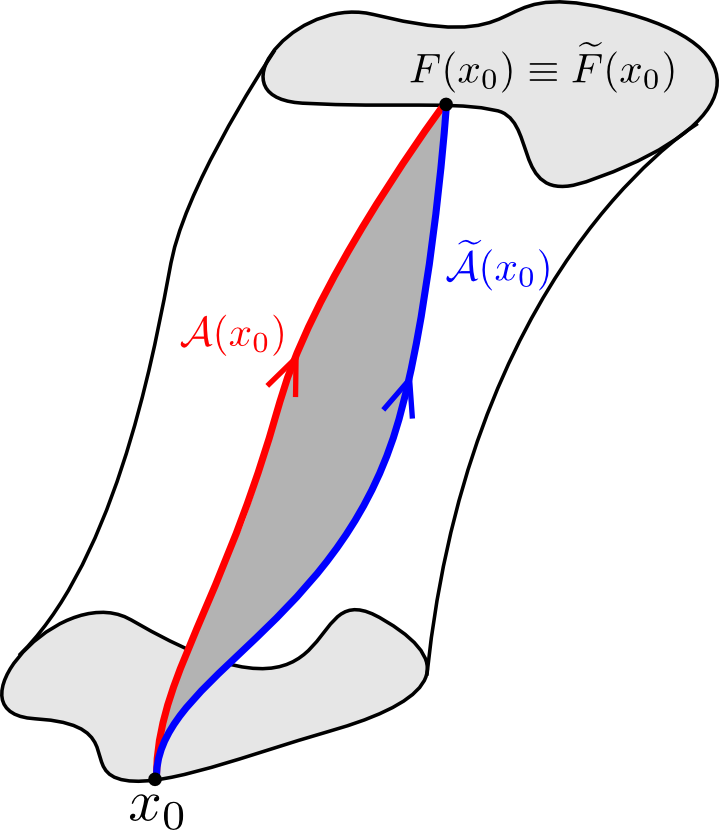}\hspace{2pc}%
\begin{minipage}[b]{18pc}\caption{\label{fig:bound}Illustration of why $\widetilde{F}=F$ implies $\widetilde{\cal A}={\cal A}$ when the winding numbers are the same. Both braids have the same ${\bf n}\times{\bf A}$ on $\partial V$, although they may have different field lines on this boundary linking $x_0$ and $F(x_0)$ (as shown). But the flux through the loop must vanish, so $\widetilde{\cal A}={\cal A}$ by Stokes' Theorem.}
\end{minipage}
\end{figure}

We already know that ${\cal A}$ is a \emph{necessary} condition for topological equivalence, because it is an ideal invariant. But we can also see this from Lemma \ref{lem:da}. Assuming $\widetilde{F}=F$, it follows that
\begin{equation}
\mathrm{d}\widetilde{\cal A} = \widetilde{F}^*\alpha-\alpha=F^*\alpha-\alpha=\mathrm{d}{\cal A},
\end{equation}
so that $\widetilde{\cal A}$ and ${\cal A}$ differ by at most an overall constant. But since $\widetilde{F}|_{\partial D_0}=F_{\partial D_0}$ (with the same winding number) we must have $\widetilde{\cal A}={\cal A}$ (see Figure \ref{fig:bound}).

\subsection*{Sufficiency}
To prove the converse---that ${\cal A}$ is a \emph{sufficient} condition for topological equivalence---assume that ${\cal A}=\widetilde{\cal A}$ and define the mapping $G=\widetilde{F}\circ F^{-1}$. Then, using Lemma \ref{lem:da},
\begin{align}
G^*\alpha-\alpha &= (F^{-1})^*\circ\widetilde{F}^*\alpha - \alpha\\
&= (F^{-1})^*(\alpha + \mathrm{d}{\cal A}) - \alpha\\
&= (F^{-1})^*\alpha - \alpha + (F^{-1})^*\mathrm{d}{\cal A}\\
&= (F^{-1})^*\alpha - (F^{-1})^*\circ F^* \alpha+ (F^{-1})^*\mathrm{d}{\cal A}\\
&= (F^{-1})^*(\alpha - F^*\alpha)+ (F^{-1})^*\mathrm{d}{\cal A}\\
&= -(F^{-1})^*\mathrm{d}{\cal A}+ (F^{-1})^*\mathrm{d}{\cal A}\\
&= 0.
\end{align}
Now we determine the possible mappings $G:D_1\rightarrow D_1$ satisfying $G^*\alpha=\alpha$, which \citet{haro1998,haro2000}
\index{authors}{Haro, A.}%
 calls ``actionmorphisms''. If $\alpha$ is in the canonical gauge $A^{\rm ref}_\rho=0$, then the possible mappings $G$ that preserve $\alpha$ are known \citep[][Proposition 6.3.2]{marsden1994} 
\index{authors}{Marsden, J.E.}%
\index{authors}{Ratiu, T.S.}%
 to take the form of cotangent lifts
\begin{equation}
G(q,p)=\left(T(q),\frac{p}{T'(q)} \right)
\label{eqn:lifts}
\end{equation}
where $T$ is a diffeomorphism. Since $\partial D_0$ is not a line of constant $q(\equiv\psi)$, it follows from our assumption $\widetilde{F}|_{\partial D_0}=F|_{\partial D_0}$ that $T(q)=q$ and hence that $G_q=q$, $G_p=p$. We see that $G_\psi=\psi$. Recalling that $p=A^{\rm ref}_\psi h_\psi$, we note in canonical gauge that $B_z>0$ implies that $\partial p/\partial\rho>0$. It follows that the coordinate transformation from $(\rho,\psi)$ to $(p,q)$ has non-zero Jacobian, and hence that $G_\rho=\rho$. So $G={\rm id}$ and therefore $\widetilde{F}=F$. \hfill \qedsymbol

\subsection*{Remarks}

\begin{enumerate}
\item If $\alpha$ is not in the canonical gauge, then $G$ need not take the form of a cotangent lift. Indeed, the nature of the actionmorphisms appears to depend on the gauge of $\alpha$. For suppose that $G^*\alpha-\alpha=0$ for some $\alpha$. Then after a gauge transformation $\alpha \rightarrow \alpha + \mathrm{d}\chi$ we find
\begin{equation}
G^*(\alpha+\mathrm{d}\chi) - (\alpha+\mathrm{d}\chi) = \mathrm{d}(G^*\chi - \chi) \neq 0
\end{equation}
so that $G$ is no longer an actionmorphism in the new gauge.\index{subject}{gauge transformation}

\item Notice that the property of whether or not two magnetic braids are topologically equivalent does not depend on the choice of reference field. This choice is arbitrary; all that matters is that the \emph{same} reference field is used to compute both ${\cal A}$ and $\widetilde{\cal A}$. On the other hand, the nature of differences in ${\cal A}$ between two inequivalent braids may depend on the choice of reference: this is an issue for future investigation.
\end{enumerate}

\section{Examples: Magnetic Braids on a Cylinder} \label{sec:ex}

We conclude with two simple examples on the cylinder $V=\{(r,\phi,z):0\leq r\leq 4, -8\leq z\leq 8 \}$. The first serves to illustrate the theory with a simple magnetic braid where ${\cal A}$ may be calculated analytically, and the second to illustrate an important topological interpretation of ${\cal A}$ introduced by \citet{yeates2012}
\index{authors}{Yeates, A.R.}%
\index{authors}{Hornig, G.}%
 for magnetic braids in the cylinder.

\subsection{Centred Flux Ring}\label{sec:ex1}

Take standard cylindrical coordinates $\rho=r$ and $\psi=\phi$, so that the scale factors are $h_\rho=h_z=1$, $h_\psi=r$. Consider the magnetic braid
\begin{equation}
{\bf B} = {\bf e}_z + \sqrt{2}r\,e^{-r^2/2 - z^2/4} {\bf e}_\phi
\label{eqn:ringb}
\end{equation}
which consists of a localised toroidal ``flux ring'' in a uniform background field \citep{wilmotsmith2009}.
\index{authors}{Wilmot-Smith, A.L.}%
\index{authors}{Hornig, G.}%
\index{authors}{Pontin, D.I.}%
 For this field the field lines are
\begin{equation}
f_r(r,\phi; z) = r,\qquad
f_\phi(r,\phi;z) = \sqrt{2\pi}\,e^{-r^2/a^2}\Big({\rm erf}(z/2) - {\rm erf}(z_0/2)\Big) + \phi,
\end{equation} 
and taking $z_0\rightarrow -\infty$, $z\rightarrow\infty$ in \eqref{eqn:ringb}, the field line mapping is
\begin{equation}
F_r(r,\phi) = r, \qquad F_\phi(r,\phi) = 2\sqrt{2\pi}\,e^{-r^2/2} + \phi.
\end{equation}
A vector potential in the appropriate gauge is
\begin{equation}
{\bf A} = \frac{r}{2}{\bf e}_\phi + \sqrt{2}\,e^{-r^2/2-z^2/4}{\bf e}_z.
\end{equation}
Since $A_z\approx 0$ on all boundaries to high accuracy, the corresponding reference is ${\bf A}^{\rm ref}=(r/2){\bf e}_\phi$, giving ${\bf B}^{\rm ref}={\bf e}_z$. One can then compute the topological flux function analytically, finding
\begin{equation}
{\cal A}(r,\phi) = \sqrt{2\pi}(r^2 + 2)\,e^{-r^2/2}.
\end{equation}
\begin{figure}
\begin{center}
\includegraphics[width=0.4\textwidth]{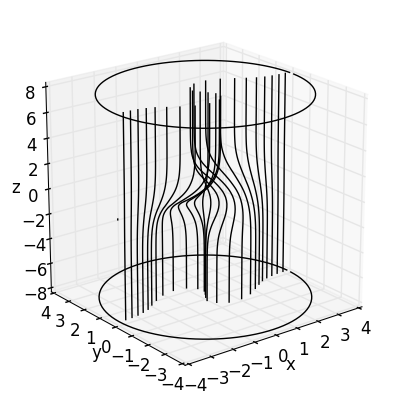}
\includegraphics[width=0.4\textwidth]{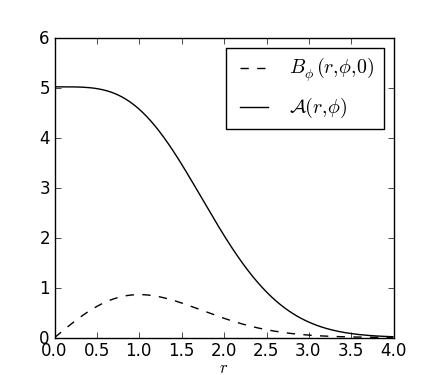}
\end{center}
\caption{\label{fig:ex}The example magnetic field from Section \ref{sec:ex1}, showing field lines (\emph{left}) and profiles of $B_\phi(z=0)$ and ${\cal A}$ as a function of $r$ (\emph{right}).}
\end{figure}
Notice that $\partial{\cal A}/\partial r = -\sqrt{2\pi}r^3\,e^{-r^2/2}$ and $\partial{\cal A}/\partial\phi = 0$.
In this case $\alpha=(r^2/2)\,\mathrm{d}\phi$ and we can verify that
\begin{equation}
\Big(F^*\alpha\Big)(r,\phi) = \frac{F_r^2}{2}\left(\frac{\partial F_\phi}{\partial r}\,\mathrm{d}r + \frac{\partial F_\phi}{\partial \phi}\,\mathrm{d}\phi\right) = \frac{r^2}{2}\left(-2\sqrt{2\pi}r\,e^{-r^2/2}\,\mathrm{d}r + \mathrm{d}\phi\right)
\end{equation}
so that indeed $F^*\alpha - \alpha = \mathrm{d}{\cal A}$.

\subsection{Displaced Flux Ring}\label{sec:ex2}

\begin{figure}
\begin{center}
\includegraphics[width=\textwidth]{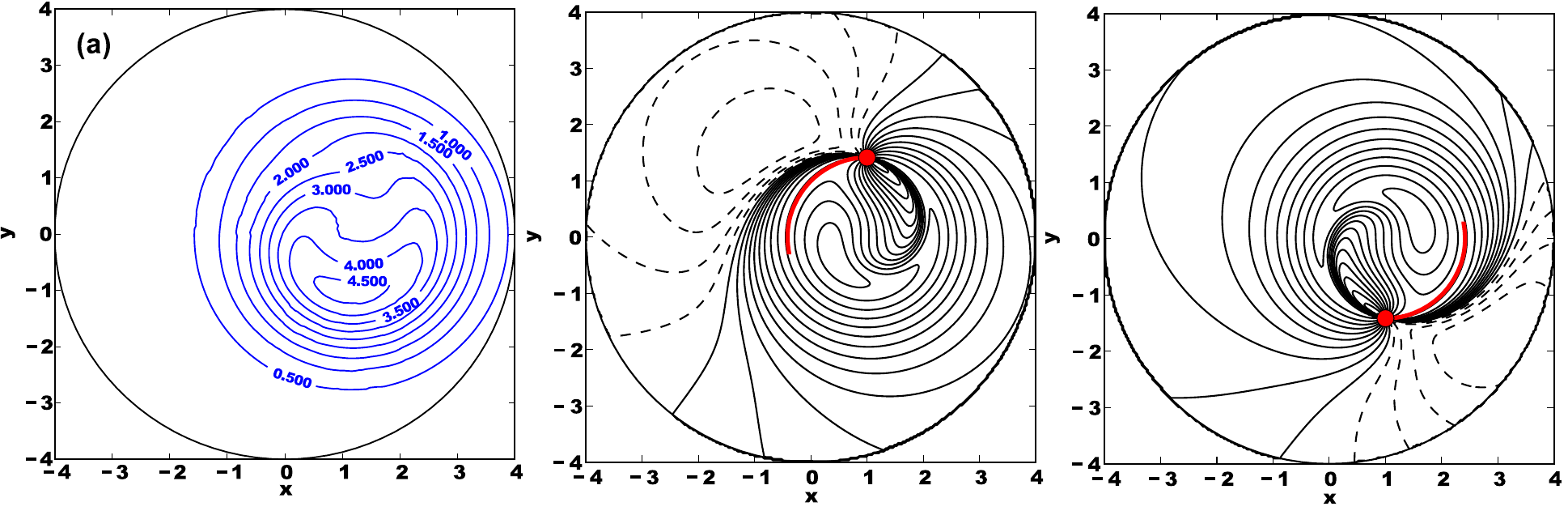}
\end{center}
\caption{\label{fig:ex2d}Asymmetry when the flux ring\index{subject}{flux ring} is displaced from the origin (Section \ref{sec:ex2}). Panel (a) shows contours of ${\cal A}$. Panels (b) and (c) show contours of the pairwise linking number $c_{x_0,y_0}$ as a function of $y_0$, for $x_0=(\sqrt{3},55^\circ)$ (in b) and $x_0=(\sqrt{3},-55^\circ)$ (in c). The red circles show $x_0$ and the thick red line shows the projection of the field line $f(x_0;z)$. Solid (dashed) contours show positive (negative) values of $c_{x_0,y_0}$. All three panels are computed numerically.
}
\end{figure}

Suppose now that the flux ring is displaced from the origin and centred at $(r=1,\phi=0)$. Then the new vector potential is
\begin{equation}
{\bf A} = \frac{r}{2}{\bf e}_\phi + \sqrt{2}\exp\left(-\frac{(r\cos\phi-1)^2 + (r\sin\phi)^2}{2}-\frac{z^2}{4} \right){\bf e}_z.
\label{eqn:move}
\end{equation}
This is not topologically equivalent to the original example, so ${\cal A}$ must differ. One might expect that the ${\cal A}$ distribution would simply be translated, but Figure \ref{fig:ex2d}(a) shows that this is not the case: the profile of ${\cal A}$ within the disturbed region is no longer symmetric about the ring centre. This reflects the fact that ${\cal A}$ is a \emph{global} quantity: it counts not just the horizontal flux of the ring itself, but how it is linked with the reference field ${\bf B}^{\rm ref}={\bf e}_z$. In \eqref{eqn:move}, the reference field is asymmetrical with respect to the position of the flux ring.

This may be understood using the following alternative formula for ${\cal A}$ on the cylinder. \citet{yeates2012}
\index{authors}{Yeates, A.R.}%
\index{authors}{Hornig, G.}%
 show that ${\cal A}(x_0)$ is the \emph{average pairwise linking number} between the field line $f(x_0;z)$ and all other field lines $f(y_0;z)$. The linking number\index{subject}{linking number} in this context \citep{berger1988}
 \index{authors}{Berger, M.A.}%
  is the integral
\begin{equation}
c_{x_0,y_0} = \frac{1}{2\pi}\int_{-8}^8\frac{\mathrm{d}\theta_{x_0,y_0}(z)}{\mathrm{d}z}\,\mathrm{d}z,
\end{equation}
where $\theta_{x_0,y_0}(z)$ is the orientation of the line connecting $f(x_0;z)$ and $f(y_0;z)$ in the plane at height $z$. It is proved by \citet{yeates2012}
\index{authors}{Yeates, A.R.}%
\index{authors}{Hornig, G.}%
 that
\begin{equation}
{\cal A}(x_0) = \int_{D_0}c_{x_0,y_0}B_z(y_0)\,\mathrm{d}^2y_0.
\label{eqn:alink}
\end{equation}
As an aside, notice that equation \eqref{eqn:alink} gives a formula to compute ${\cal A}$ directly from ${\bf B}$ without needing to know ${\bf A}$. 

The asymmetry in Figure \ref{fig:ex2d}(a) arises due to differing $c_{x_0,y_0}$ for different
field lines $f(x_0; z)$ through the flux ring. To see this, consider two field lines
$f(x_0; z)$ passing through the ring, but starting on opposite sides (Figures \ref{fig:ex2d}b and
\ref{fig:ex2d}c). The contours show $c_{x_0,y_0}$ as a function of $y_0$ for each
of these $x_0$. For $y_0$ within the ring itself, the $c_{x_0,y_0}$ contours are the same in
each case up to rotation, so these points make the same contribution to the integral \eqref{eqn:alink}. But for most $y_0$ outside the ring, $c_{x_0,y_0}$ has the opposite sign in Figure \ref{fig:ex2d}(b) compared to Figure \ref{fig:ex2d}(c), because of the field line direction $f_y(x_0;z)<0$ in Figure \ref{fig:ex2d}(b) while $f_y(x_0;z)>0$ in Figure \ref{fig:ex2d}(c). Furthermore, since $c_{x_0,y_0}$ is opposite in sign on the left and right of the ring, and since there is a larger area to the left of the ring, the net contributions to \eqref{eqn:alink} from $y_0$ outside the ring differ in each case. This explains the asymmetry. In Section \ref{sec:ex1}, the area to the left and right of the ring was the same, hence the symmetric ${\cal A}$ profile.

\ack
The work was supported by STFC Grant ST/G002436/1 to the University of Dundee. We thank J. B. Taylor, J.-J. Aly and P. Boyland for useful suggestions.

%

\bibliographystyle{jfm2}

\bibliography{Yeates_AR}

\begin{thebibliography}{22}
\expandafter\ifx\csname natexlab\endcsname\relax\def\natexlab#1{#1}\fi

\bibitem[Berger(1986)]{berger1986}
{\sc Berger, M.~A.} 1986 Topological invariants of field lines rooted to
  planes. {\em Geophys. Astrophys. Fluid Dyn.\/} {\bf 34}, 256--281.

\bibitem[Berger(1988)]{berger1988}
{\sc Berger, M.~A.} 1988 An energy formula for nonlinear force-free magnetic
  fields. {\em Astron. Astrophys.\/} {\bf 201}, 355--361.

\bibitem[Berger \& Field(1984)]{berger1984}
{\sc Berger, M.~A. \& Field, G.~B.} 1984 The topological properties of magnetic
  helicity. {\em J. Fluid Mech.\/} {\bf 147}, 133--148.

\bibitem[Berger \& Prior(2006)]{berger2006}
{\sc Berger, M.~A. \& Prior, C.} 2006 The writhe of open and closed curves.
  {\em J. Phys. A: Math. Theor.\/} {\bf 39}, 8321--8348.

\bibitem[Boozer(1983)]{boozer1983}
{\sc Boozer, A.~H.} 1983 Evaluation of the structure of ergodic fields. {\em
  Phys. Fluids\/} {\bf 26}, 1288--1291.

\bibitem[Brown {\em et~al.\/}(1999)Brown, Canfield \& Pevtsov]{brown1999}
{\sc Brown, M.~R., Canfield, R.~C. \& Pevtsov, A.~A.} (ed.) 1999 {\bf Magnetic
  Helicity in Space and Laboratory Plasmas}, Washington. AGU.

\bibitem[Candelaresi \& Brandenburg(2011)]{candelaresi2011}
{\sc Candelaresi, S. \& Brandenburg, A.} 2011 Decay of helical and nonhelical
  magnetic knots. {\em Phys. Rev. E\/} {\bf 84}, 01646.

\bibitem[Cary \& Littlejohn(1983)]{cary1983}
{\sc Cary, J.~R. \& Littlejohn, R.~G.} 1983 Noncanonical hamiltonian mechanics
  and its application to magnetic field line flow. {\em Annal. Phys.\/} {\bf
  151}, 1--34.

\bibitem[Finn \& Antonsen(1985)]{finn1985}
{\sc Finn, J.~H. \& Antonsen, T.~M.} 1985 Magnetic helicity: What is it and
  what is it good for? {\em Comments Plasma Phys. Contr. Fusion\/} {\bf 9},
  111--126.

\bibitem[Frankel(1997)]{frankel1997}
{\sc Frankel, T.} 1997 {\bf The Geometry of Physics}. CUP, Cambridge.

\bibitem[Haro(1998)]{haro1998}
{\sc Haro, A.} 1998 The primitive function of an exact symplectomorphism. PhD
  thesis, Universitat de Barcelona, Barcelona.

\bibitem[Haro(2000)]{haro2000}
{\sc Haro, A.} 2000 The primitive function of an exact symplectomorphism. {\em
  Nonlinearity\/} {\bf 13}, 1483--1500.

\bibitem[Marsden \& Ratiu(1994)]{marsden1994}
{\sc Marsden, J.~E. \& Ratiu, T.~S.} 1994 {\bf Introduction to Mechanics and
  Symmetry}. Springer-Verlag, New York.

\bibitem[Morrison(2000)]{morrison2000}
{\sc Morrison, P.~J.} 2000 Magnetic field lines, hamiltonian dynamics, and
  nontwist systems. {\em Phys. Plasmas\/} {\bf 7}, 2279--2289.

\bibitem[Reale(2010)]{reale2010}
{\sc Reale, F.} 2010 Coronal loops: Observations and modeling of confined
  plasma. {\em Living Rev. Solar Phys.\/} {\bf 7}, 5.

\bibitem[Taylor(1986)]{taylor1986}
{\sc Taylor, J.~B.} 1986 Relaxation and magnetic reconnection in plasmas. {\em
  Rev. Mod. Phys.\/} {\bf 58}, 741--763.

\bibitem[Wilmot-Smith {\em et~al.\/}(2009)Wilmot-Smith, Hornig \&
  Pontin]{wilmotsmith2009}
{\sc Wilmot-Smith, A.~L., Hornig, G. \& Pontin, D.~I.} 2009 Magnetic braiding
  and parallel electric fields. {\em Astrophys. J.\/} {\bf 696}, 1339--1347.

\bibitem[Woltjer(1958)]{woltjer1958}
{\sc Woltjer, L.} 1958 A theorem on force-free magnetic fields. {\em Proc. Nat.
  Acad. Sci.\/} {\bf 44}, 489--491.

\bibitem[Yeates \& Hornig(2011)]{yeates2011}
{\sc Yeates, A.~R. \& Hornig, G.} 2011 A generalized flux function for
  three-dimensional magnetic reconnection. {\em Phys. Plasmas\/} {\bf 18},
  102118.

\bibitem[Yeates \& Hornig(2013)]{yeates2012}
{\sc Yeates, A.~R. \& Hornig, G.} 2013 Unique topological characterization of
  braided magnetic fields. {\em Phys. Plasmas\/} {\bf 20}, 012102.

\bibitem[Yeates {\em et~al.\/}(2010)Yeates, Hornig \& Wilmot-Smith]{yeates2010}
{\sc Yeates, A.~R., Hornig, G. \& Wilmot-Smith, A.~L.} 2010 Topological
  constraints on magnetic relaxation. {\em Phys. Rev. Lett.\/} {\bf 105},
  085002.

\bibitem[Yoshida(1994)]{yoshida1994}
{\sc Yoshida, Z.} 1994 A remark on the hamiltonian form of the
  magnetic-field-line equations. {\em Phys. Plasmas\/} {\bf 1}, 208--209.

\end{thebibliography}
\end{document}